\documentclass{elsart1p}
\usepackage{graphicx}
\usepackage{amssymb}
\def\gtap{\ \raise.3ex\hbox{$>$\kern-.75em\lower1ex\hbox{$\sim$}}\ }
\begin{document}
\begin{frontmatter}
%
%
%
\title{Coherent Pion Production in the Neutrino-Nucleus Scattering in Few-GeV
Region}
%
%
\author[triumf]{S. X. Nakamura\corauthref{cor1}}
\author[osaka]{T. Sato}
\author[anl]{T.-S. H. Lee}
\author[sd]{B. Szczerbinska}
\author[usc]{K. Kubodera}
\address[triumf]{TRIUMF, 4004 Wesbrook Mall, Vancouver, BC, V6T 2A3, Canada}
\address[osaka]{Department of Physics, Osaka University, Toyonaka, Osaka,
 560-0043, Japan}
\address[anl]{Physics Division, Argonne National Laboratory, Argonne, Illinois 60439, USA}
\address[sd]{Dakota State University, College of Arts \& Sciences,
Madison, SD, 57042-1799, USA}
\address[usc]{Department of Physics and Astronomy, University of South
 Carolina, Columbia, SC, 29208, USA}
\corauth[cor1]{snakamura@triumf.ca}
\begin{abstract}
We study coherent pion production in neutrino-nucleus scattering in
the energy region relevant to the recent neutrino oscillation experiments.
Our approach is based on the combined use of the Sato-Lee model 
and the $\Delta$-hole model.
Our initial numerical results are compared with 
the recent data from K2K and SciBooNE.

\end{abstract}
\begin{keyword}
neutrino-nucleus scattering \sep neutrino oscillation \sep pion production
%
\PACS 13.15.+g \sep 14.60.Pq \sep 25.30.Pt
\end{keyword}
\end{frontmatter}
%
\section{Introduction}
\label{intro}

Recent neutrino oscillation experiments, such as T2K and MiniBooNE,
point to the urgency of more theoretical work on
neutrino-nucleus reactions in the sub- and few-GeV region.  
One of such reactions is coherent pion production 
in which the final nucleus remains in its ground state. 
The processes in neutrino-nucleus scattering
are dominated by quasi-elastic scattering and 
quasi-free $\Delta$-excitation (with its subsequent
decay), and the coherent processes represent a small fraction.
Even so, it is important to take accurate account 
of the coherent processes in order to achieve 
a significant reduction of theoretical uncertainties
in interpreting data of the neutrino oscillation experiments.
Lately, several neutrino experimental groups presented
new data on coherent pion production in both
charged-current (CC)\cite{hasegawa,hiraide} and neutral-current (NC)
reactions\cite{miniboone}, and also for the anti-neutrino
reaction\cite{anti}.
It is therefore timely to develop a theoretical model 
whose accuracy matches the precision of these new data.
We report here our study of coherent pion production in
neutrino-nucleus scattering, explaining a calculational framework
we have recently developed and presenting numerical results.

\section{Model}
\label{model}

As a starting point for studying pion production in neutrino-nucleus
scattering, we need a model which reasonably describes
electroweak pion production off a single nucleon. 
We employ here the Sato-Lee (SL) model\cite{SL}, 
which is considered to provide a reliable framework for this purpose.
In applying the SL model to a reaction on a nuclear target,
we need to consider medium effects as well. 
Major medium effects are known to be the change of 
the $\Delta$-propagation (shifts of the mass and width) 
and the final-state interactions (FSI) between 
the outgoing pion and nucleus. 
In the energy region of our interest, 
the $\Delta$-hole model\cite{karaoglu} is believed
to allow us to incorporate these medium effects. 
We therefore expect 
that the combined use of the SL model
and the $\Delta$-hole model
will lead to a reasonable theoretical framework 
for calculating pion production in nuclei. 

We proceed as follows. 
First, we construct an optical potential for
pion-nucleus scattering; 
this potential is used to take account of the FSI
on the outgoing pion.
Our model involves four free complex parameters; two of them are
responsible for the medium effect on the $\Delta$
(the so-called spreading potential), 
and the other two pertain to phenomenological
pion-nucleus interactions, 
which are proportional to the square of the
nuclear density.
All of these parameters are fixed so as to optimize fit 
to experimental data for pion-nucleus scattering 
(both elastic and total cross sections).
We then are in a position to calculate 
the coherent pion production process 
without free parameters.
To test the reliability of our approach, 
we apply our model to coherent pion ``photo''-production 
and compare the results with data.
After this test, we calculate 
coherent pion production in neutrino-nucleus scattering.
The results are compared with the recent data from K2K\cite{hasegawa}.

\section{Results}
\label{result}

\begin{figure}[t]
\begin{minipage}[t]{65mm}
\includegraphics[width=62mm]{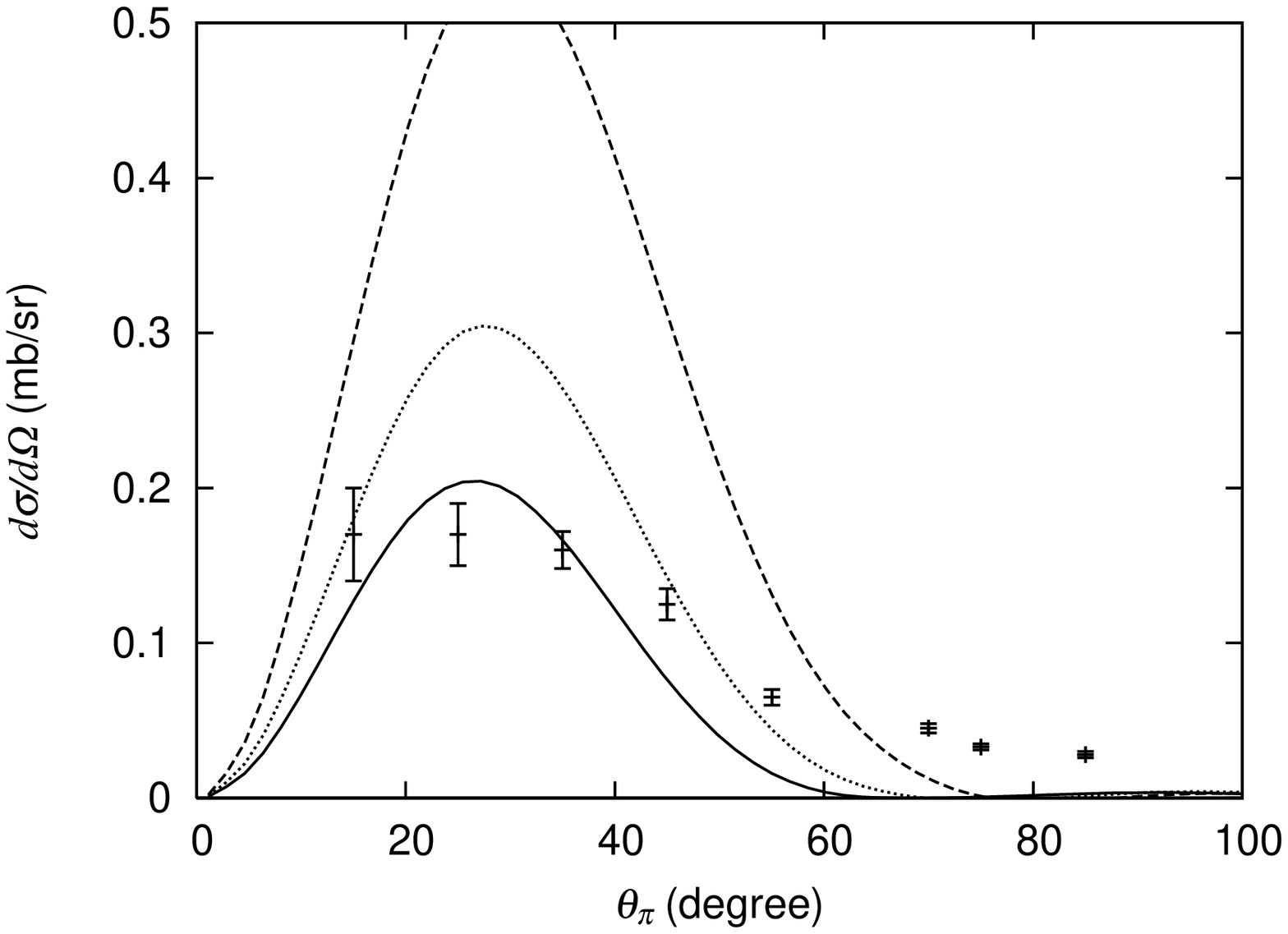}
 \caption{\label{fig1}
Angular distribution of the pion for
$\gamma + {}^{12}{\rm C}\to \pi^0 + {}^{12}{\rm C}_{g.s.}$
($E_\gamma = 290$~MeV); laboratory frame. 
For explanation of the curves, see the text.
The data are from Ref.~\cite{zpa311}.
}
\end{minipage}      
\hspace{3mm}
\begin{minipage}[t]{65mm}
 \includegraphics[width=62mm]{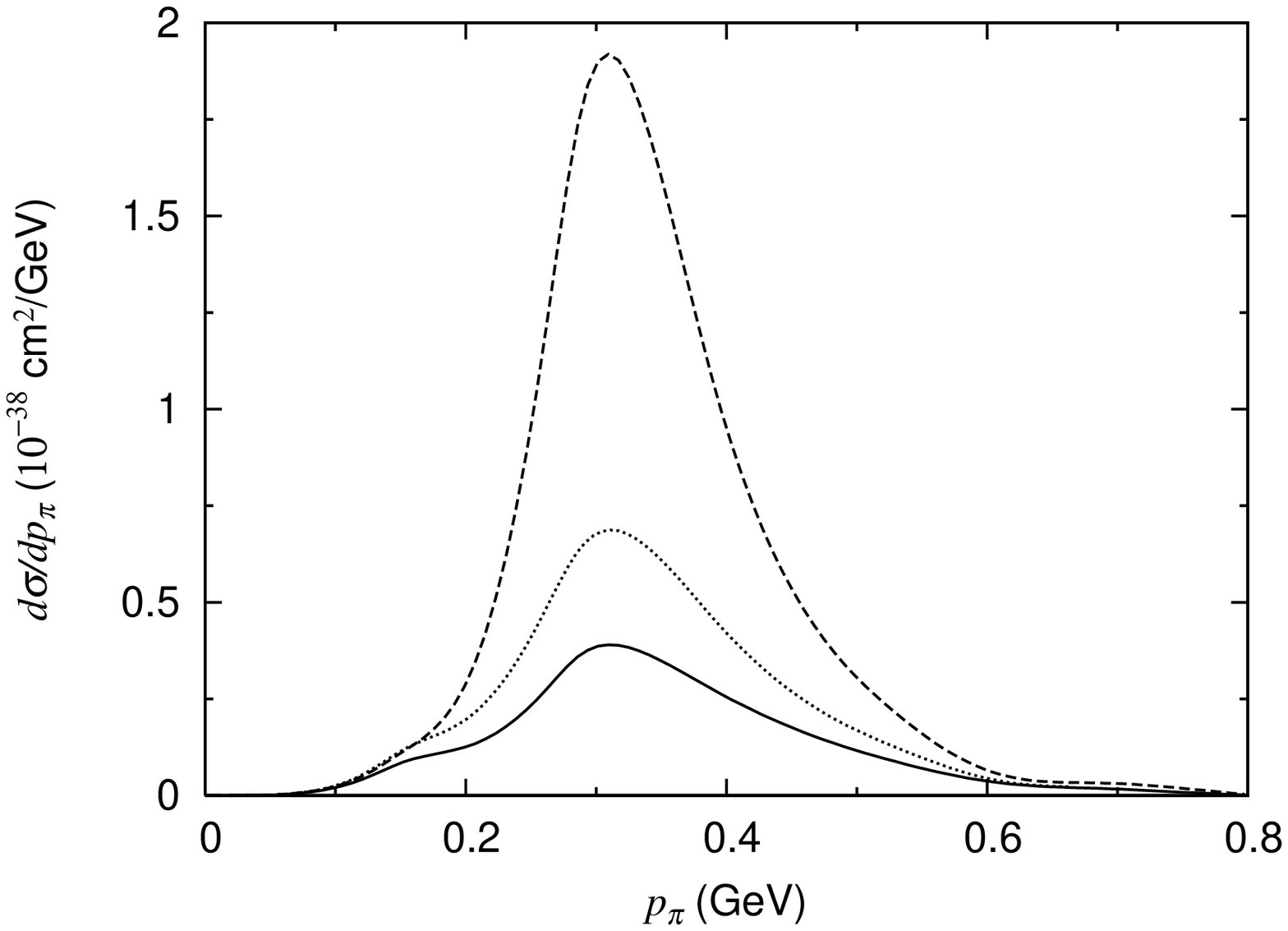}
\caption{\label{fig2}Pion momentum distribution for 
$\nu_\mu + {}^{12}{\rm C}\to \mu^- + \pi^+ +  {}^{12}{\rm C}_{g.s.}$
($E_\nu = 1$~GeV); neutrino-nucleus CM frame.
For explanation of the curves, see the text.
}
\end{minipage}
\end{figure}

Figure~\ref{fig1} shows our results 
for the pion angular distribution  
in coherent pion photo-production off $^{12}$C. 
The dashed curve represents the case
that does not include FSI
or the shifts of the $\Delta$ mass and width;
medium effects enter 
only through the nuclear form factor here.
The results obtained with FSI included are given
by the dotted curve, 
while the results of our full calculation
that takes into account both FSI and the in-medium
modification of the $\Delta$ properties 
are shown by the solid curve.
Our results are fairly consistent with the data around the peak. 
The large discrepancy between the calculation and the data at
$\theta_\pi\gtap 60^\circ$ is due to the fact that the data include
incoherent contributions from low-lying excited states.
The reasonable agreement between theory and experiment 
seen in Fig.~\ref{fig1} provides a good basis to proceed
to calculate the neutrino-induced process. 

Our results for neutrino-induced CC coherent pion
production are displayed in Fig.~\ref{fig2}; 
the figure shows the pion momentum spectrum 
for $E_\nu =$ 1~GeV ($E_\nu$: neutrino energy). 
Here again, 
the dashed curve corresponds to the case
in which FSI and the shifts of 
the $\Delta$ mass and width are ignored,
the dotted curve to the case that only includes FSI,
and the solid curve to the full calculation that
includes FSI and the in-medium
modification of the $\Delta$ properties.
The important role of the medium effects is clearly seen.
The total cross section is 
9.5 $\times 10^{-40} {\rm cm}^2$ for
$E_\nu =$ 1~GeV.
In Ref.~\cite{hasegawa}, the upper limit of 
the total cross section averaged over 
the neutrino spectrum is given; 
$7.7\times 10^{-40} {\rm cm}^2$. 
(Reference \cite{hiraide} presented a
similar result.)
Although the total cross section averaged 
over the neutrino spectrum 
is yet to be calculated,
it seems that our results are significantly larger than
the experimental upper limit.
It is noted, however, that the upper limit 
was obtained from the analysis which was
partly based on theoretical models (quasi-elastic, resonant 1$\pi$, etc.)
which are still under active debate.
A better understanding of the other processes together with the pion
coherent production process is highly desirable.

\section{Summary}
\label{summary}

In this work we constructed a model for
coherent pion production in neutrino-nucleus scattering
around 1~GeV.
The model is based on a combination of the SL model and the
$\Delta$-hole model.
We fixed all the free parameters in our model 
by fitting to pion-nucleus scattering data.
The results for the photon-induced coherent pion production were found
to be in reasonable agreement with the data.
For the neutrino-induced process, however, 
our initial results are not consistent 
with the recent data from K2K and SciBooNE.
We plan to study the origin of this inconsistency, and then to
apply our model to the NC reaction and the anti-neutrino
process.

\vspace{5mm}
\noindent {\bf \large Acknowledgment}

\noindent This work is supported by
the Natural Sciences and Engineering Research Council of Canada(SXN),
by the U.S. Department of Energy, Office of Nuclear Physics,
under contract DE-AC02-06CH11357 (TSHL),
by the Japan Society for the Promotion of Science,
Grant-in-Aid for Scientific Research(C) 20540270 (TS),
and by the U.S. National Science Foundation
under contract PHY-0758114 (KK).

%
%
%

%

\begin{thebibliography}{00}
%
%
%
%
\bibitem{hasegawa} M. Hasegawa {\it et al.},
Phys. Rev. Lett. {\bf 95}, 252301 (2005).

\bibitem{hiraide} K. Hiraide {\it et al.}, arXiv:0811.0369 [hep-ex].

\bibitem{miniboone}
A.A. Aguilar-Arevalo {\it et al.}, Phys. Lett. B {\bf 664}, 41 (2008). 

\bibitem{anti}  V.T. Nguyen, arXiv:0806.2347 [hep-ex].


\bibitem{SL} T. Sato and T.-S. H. Lee, Phys. Rev. C {\bf 54}, 2660 (1996);\\
   T. Sato, D. Uno and T.-S. H. Lee, Phys. Rev. C {\bf 67}, 065201 (2003).

\bibitem{karaoglu} B. Karaoglu and E. J. Moniz, Phys. Rev. C {\bf 33}, 974 (1986).


\bibitem{zpa311} J. Arends {\it et al.}, Z. Phys. A {\bf 311}, 367 (1983).

%
\end{thebibliography}
\end{document}